\documentclass[article]{revtex4}

\usepackage{psfig}
\usepackage{graphicx}%
\usepackage{dcolumn}
\usepackage{amsmath}

\makeatletter
\def\btt#1{\texttt{\@backslashchar#1}}%
\DeclareRobustCommand\bblash{\btt{\@backslashchar}}%
\makeatother

\begin{document}

\title{Reconstructing the Equation of State of Tachyon}
\
\author{Jian-gang Hao}

\author{Xin-zhou Li}\email{kychz@shtu.edu.cn}

\affiliation{ Shanghai United Center for Astrophysics, Shanghai
Normal University, Shanghai 200234 , China
}%

\date{\today}

\begin{abstract}
Abstract : Recent progress in theoretical physics suggests that
the dark energy in the universe might be resulted from the rolling
tachyon field of string theory. Measurements to SNe Ia can be
helpful to reconstruct the equation of state of the rolling
tachyon which is a possible candidate of dark energy. We present a
numerical analysis for the evolution of the equation of state of
the rolling tachyon and derive the reconstruction equations for
the equation of state as well as the potential.
\end{abstract}


\maketitle

\vspace{0.4cm} \noindent\textbf{I. Introduction} \vspace{0.4cm}

In the evolution model of the universe, the underlying dynamics is
mostly described as a single scalar field rolling in certain
potential. This scenario is preferred because of its simplicity
and catches the most attention since its proposition. Furthermore,
many seemingly more complicated models can be rewritten in such a
framework. An important feature of the spectra of density
perturbation and gravitational perturbation, which has been widely
studied in inflation theory, is that they can be linked by a
consistency euqation. Given a particular set of observations of
certain accuracy, one may attempt the bold task of reconstructing
the potential of the scalar field from the observations. The study
on the reconstruction equation was firstly introduced when
investigating the inflation universe\cite{turner1}, in which it
was usually referred to as "perturbation reconstruction". In this
scheme, One has found the relationship between observations of
microwave anisotropies and that of the large-scale structure, and
tried to connecting them with the potential of the scalar field
that drives the inflation. In this approach, the consistency
equation and scalar potential are determined as an expansion about
a given point( regarded either as a single scale in the spectra or
as a single point on the potential), allowing the reconstruction
of a region of the potential about the point. The second round of
study on reconstruction equations appears in the investigation of
quintessence\cite{turner2, steinhardt}, in which the connection
between the red-shift of the supernovae and the potential of the
quintessence is established.

In this paper, following the ingenuity of the above
reconstruction, we consider the reconstruction of the equation of
state as well as the potential of the rolling tachyon, which has
been proposed recently to play the role of dilaton \cite{Sen,
Gibbons, Fairbairn, Feinstein, Mukohyama, Gerasimov, Mazumdar,
Padmanabhan, Frolov, Minahan, Kim, Choudhury, shiu, Linde,
Maggiore, Sugimoto,Benaoum, Sami, Li1} or quintessence\cite{Li}.
The potential of the rolling tachyon could be derived from string
theory and its dynamics has been widely studied. On the other
hand, one can also reconstruct the effective potential of the
tachyon field by the observations. It should be noted that the CMB
anistropic spectra mainly probe the universe at redshift $z \sim
1000$ when the ratio of dark energy to matter is less than
$10^{-6}$. While the Supernovae observations reflect that the
universe at recent epoch when the dark energy is beginning to
dominate the universe. Therefore, if we assume that the rolling
tachyon is the candidate for dark energy, we'd better reconstruct
the potential of the tachyon field by the red-shifts data of SNe
Ia.

It is worth noting that the effective potential of the tachyon
field should be derived from sting theory by considering the
corresponding process. While it can also be reconstructed from the
observations. It would be more appropriate to consider the
reconstruction equation as a criteria with which one can judge
whether the effective potential from string theory can fit the SNe
Ia red-shift data. The outline of this paper is as follows: In
section II, we exhibit the reconstruction of the potential of
tachyon field. Section III contains a brief numerical analysis for
the dynamical evolution of tachyon and the conclusions are
discussed in section IV.

\vspace{0.4cm} \noindent\textbf{II. Reconstruct the Potential of
Tachyon field}
 \vspace{0.4cm}

The effective Lagrangian density of tachyon of string theory in a
flat Robertson-Walker background is as following:
\begin{equation}
L=-V(T)\sqrt{1+g^{\mu\nu}\partial_\mu T\partial_\nu T}
\end{equation}

\noindent where
\begin{equation}
ds^{2}=-dt^{2}+a^{2}(t)(dx^2+dy^2+dz^2)
\end{equation}

\noindent is the flat Robertson-Walker metric and $V(T)$ is the
potential resulted from string theory. When we consider the
existence of non-relativistic matter and the tachyon field $T$,
the Einstein equations for the evolution of the background
metric,$G_{\mu\nu}=\kappa T_{\mu\nu}$ can be written as:

\begin{equation}\label{H}
H^2=(\frac{\dot{a}}{a})^2=\frac{\kappa}{3}(\rho_T+\rho_M)
\end{equation}
\noindent and
\begin{equation}\label{dH}
\frac{\ddot{a}}{a}=-\frac{\kappa}{3}(\frac{1}{2}\rho_T +
\frac{3}{2}p_T+\frac{1}{2}\rho_M)
\end{equation}

\noindent For a spatially homogenous tachyon field $T$, we have
the equation of motion
\begin{equation}\label{T}
\ddot{T}
+3H\dot{T}(1-\dot{T}^2)+\frac{V^{'}(T)}{V(T)}(1-\dot{T}^2)=0
\end{equation}

\noindent where the overdot represents the differentiation with
respect to $t$ and the prime denotes the differentiation with
respect to $T$. Eq.(\ref{T}) is also equivalent to the entropy
conservation equation. The constant $\kappa=8\pi G$ where $G$ is
Newtonian gravitation constant. The density $\rho_T$ and the
pressure $p_T$ are defined as following:

\begin{equation}
\rho_T=\frac{V(T)}{\sqrt{1-\dot{T}^2}}
\end{equation}

\begin{equation}
p_T=-V(T)\sqrt{1-\dot{T}^2}
\end{equation}

\noindent the equation of state is

\begin{equation}
w_T=\frac{p_T}{\rho_T}=\dot{T}^2-1
\end{equation}

It is clear that when tachyon field is dominant and if the tachyon
field can accelerate the expansion of the universe, there must be
$\dot{T}^2<\frac{2}{3}$ and $-1<w_T<-\frac{1}{3}$.

Now, we will correlate the potential with the observable red-shift
of SNe Ia. To do so, following the earlier study in this
field\cite{turner2}, we introduce the quantity
\begin{equation}
  r(z)=\int^{t_0}_{t(z)}\frac{du}{a(u)}=\int^z_0\frac{dx}{H(x)}
\end{equation}

\noindent which is the Robertson-Walker coordinate distance to an
object at red-shift $z$. Also, we denote
\begin{equation}\label{rhom}
\rho_M=\Omega_M \rho_{crit}=\frac{3\Omega_M H_0^2(1+z)^3}{\kappa}
\end{equation}

\noindent where $H_0$ is the present Hubble constant, $\Omega_M$
is the fraction of non-relativistic matter to the critical density
and $\rho_{crit}$ is the critical energy density of the universe.

We then readily have
\begin{equation}
  (\frac{\dot{a}}{a})^2=H(z)^2=\frac{1}{(dr/dz)^2}
\end{equation}

\begin{equation}
 \frac{\ddot{a}}{a}=\frac{1}{(dr/dz)^2}+(1+z)\frac{d^2r/dz^2}{(dr/dz)^3}
\end{equation}

\begin{equation}\label{relation}
  \frac{dz}{dt}=-(1+z)H(z)=-(1+z) \frac{dr}{dz}
\end{equation}

Through Eq.(\ref{H}) to Eq.(\ref{relation}), one can express the
potential of the rolling tachyon in term of the red-shift $z$,
$dr/dz$ and $d^2r/dz^2$ as:

\begin{eqnarray}\label{zV}
  V^2(T(z))=&&[\frac{3}{\kappa (dr/dz)^2}-\rho_M ]\\\nonumber
  &&\times[\frac{3}{\kappa
(dr/dz)^2}+\frac{2(1+z)(d^2r/dz^2)}{(dr/dz)^3}]
\end{eqnarray}

\begin{eqnarray}\label{zT}
 (\frac{dT}{dz})^2=\frac{(dr/dz)^2}{(1+z)^2}-
 \frac{\kappa^2 (dr/dz)^6V(T(z))^2}{(1+z)^2[3-\kappa \rho_M(dr/dz)^2]^2}
\end{eqnarray}

\begin{equation}\label{w}
  w_T=-\frac{\kappa^2 (dr/dz)^4 V(T(z))^2}{[3-\kappa \rho_M(dr/dz)^2]^2}
\end{equation}

It is clear that if we know the coordinate distance as a function
of the red-shift $z$, which can be attained by fitting the
observation data , we can reconstruct the quantities such as
potential and the equation-of-state by the above reconstruction
equations. Also, it must be pointed out that there are sign
ambiguities in Eq.(\ref{zV}) and Eq.(\ref{zT}), which suggests
that the reconstruction equations can not completely determine the
potential as well as the variation of the field. However, the
equation of state $w_T$ can be determined uniquely by the
reconstruction equation (\ref{w}). Especially, from Eq.(\ref{w}),
one can find that $w_T$ is always negative no matter what is the
form of the $r(z)$ and $V(T(z))$. This is compatible with the
property of dark matter which possesses a negative pressure.

\vspace{0.4cm} \noindent\textbf{3. Numerical Analysis for the
Dynamical Evolution of Tachyon }
 \vspace{0.4cm}

The evolution of the tachyon is determined by its potential, which
may depend on the underlying (bosonic or supersymmetric) string
field theory. In this section, we analyze the evolution of the
equation of state $w_T$ of the rolling tachyon in the
potential\cite{gerasimov,kutasov}
\begin{equation}\label{v}
  V(T)=V_0(1+\frac{T}{T_0})\exp(-\frac{T}{T_0})
\end{equation}

Here, we consider the case that tachyon field is dominant over the
non-relativistic matter. In an earlier paper, we have shown, with
the aid of phase-plane analysis, that there is no stable critical
point in the evolution of the tachyon field\cite{Li}. The critical
point is a saddle point, which implies the evolution of the
tachyon field is very sensitive to its initial condition. Rescale
the tachyon field and the time variable by setting $T=xT_0$ and
$t=sT_0$. Introducing a new dimensionless variable
$y=\frac{dT}{dt}$, one can reduce the equation of motion of the
tachyon field in the potential Eq.(\ref{v}) to two first order
differential equations as following:

\begin{eqnarray}\label{phase}
\frac{dx}{ds}=&&y\\
\frac{dy}{ds}=&&\frac{x(1-y^2)}{1+x}-\beta
y(1-y^2)^{\frac{3}{4}}(1+x)^{\frac{1}{2}}\exp(-\frac{x}{2})
\end{eqnarray}

\noindent where $\beta=\sqrt{3V_0\kappa} T_0$ is a dimensionless
parameter. Now, it is straightforward to carry out numerical
analysis on the above equations and the following are the
numerical results (we choose $\beta$ as 0.9,1.0 and 1.1
respectively) .

\begin{figure}
\centerline{\psfig{figure=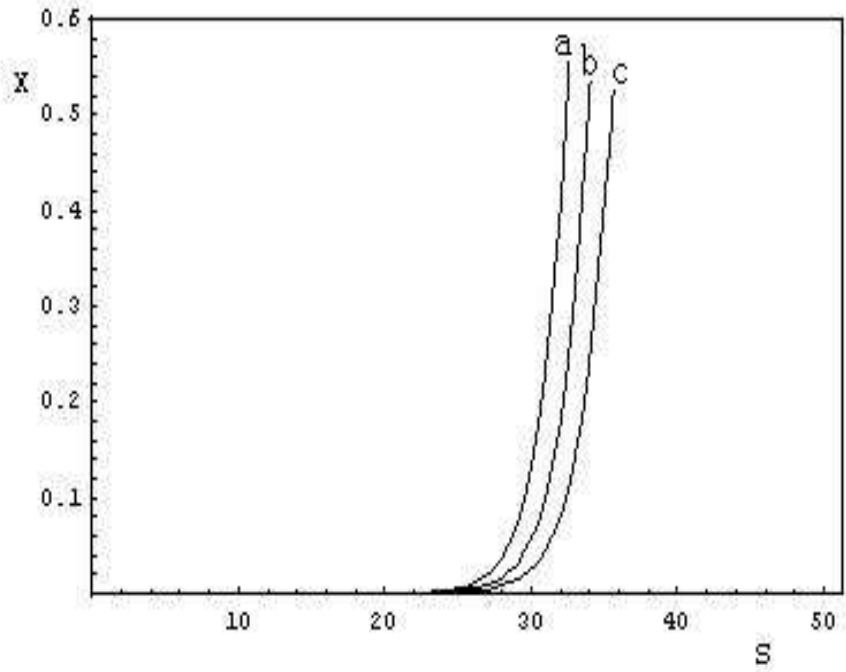,width=6in,height=4in}}
\caption{The plot of $x$ against $s$, curve a, b and c correspond
to the cases of $\beta=0.9, 1.0, 1.1$ respectively.}
\end{figure}

\begin{figure}
\centerline{\psfig{figure=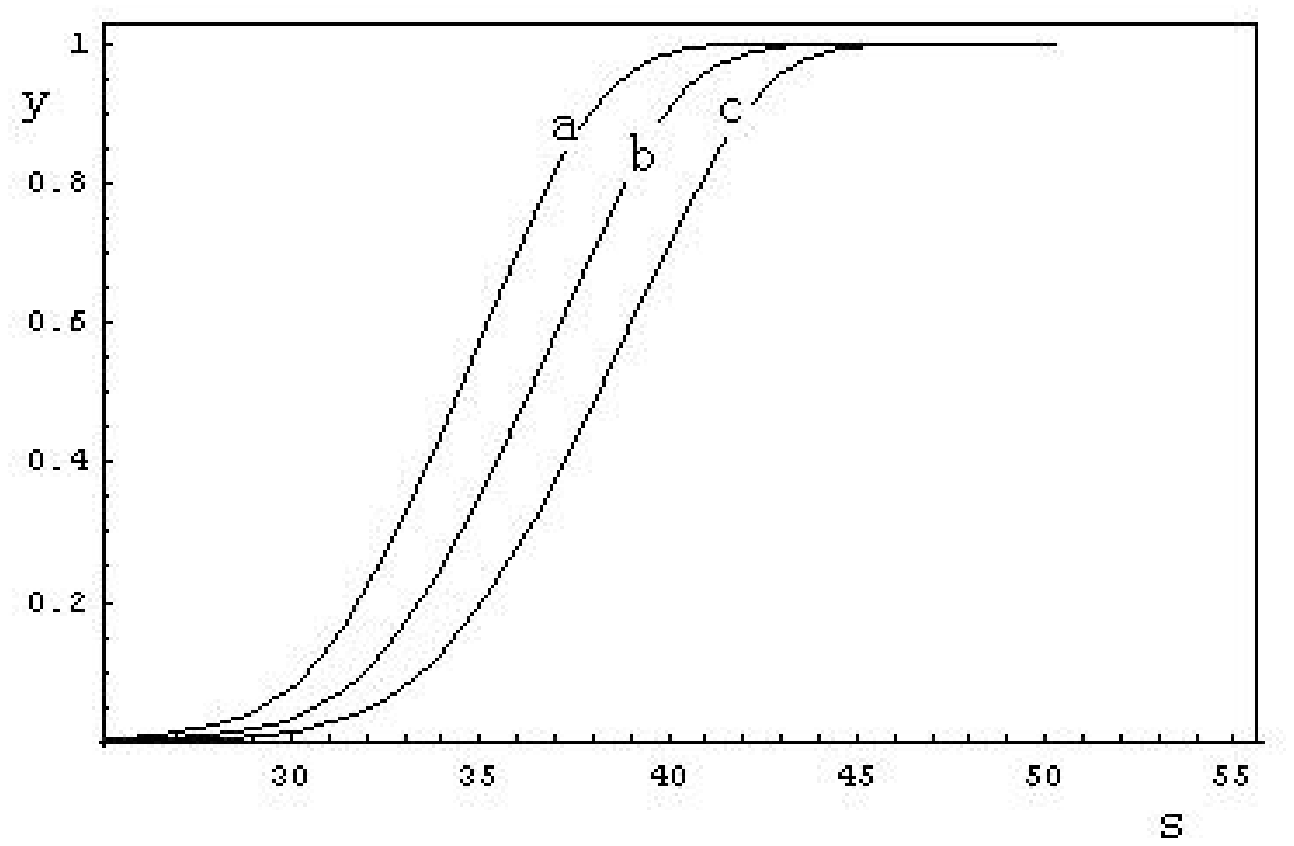,width=6in,height=4in}}
\caption{The plot of $y$ against $s$, curve a, b and c correspond
to the cases of $\beta=0.9, 1.0, 1.1$ respectively.}
\end{figure}

\begin{figure}
\centerline{\psfig{figure=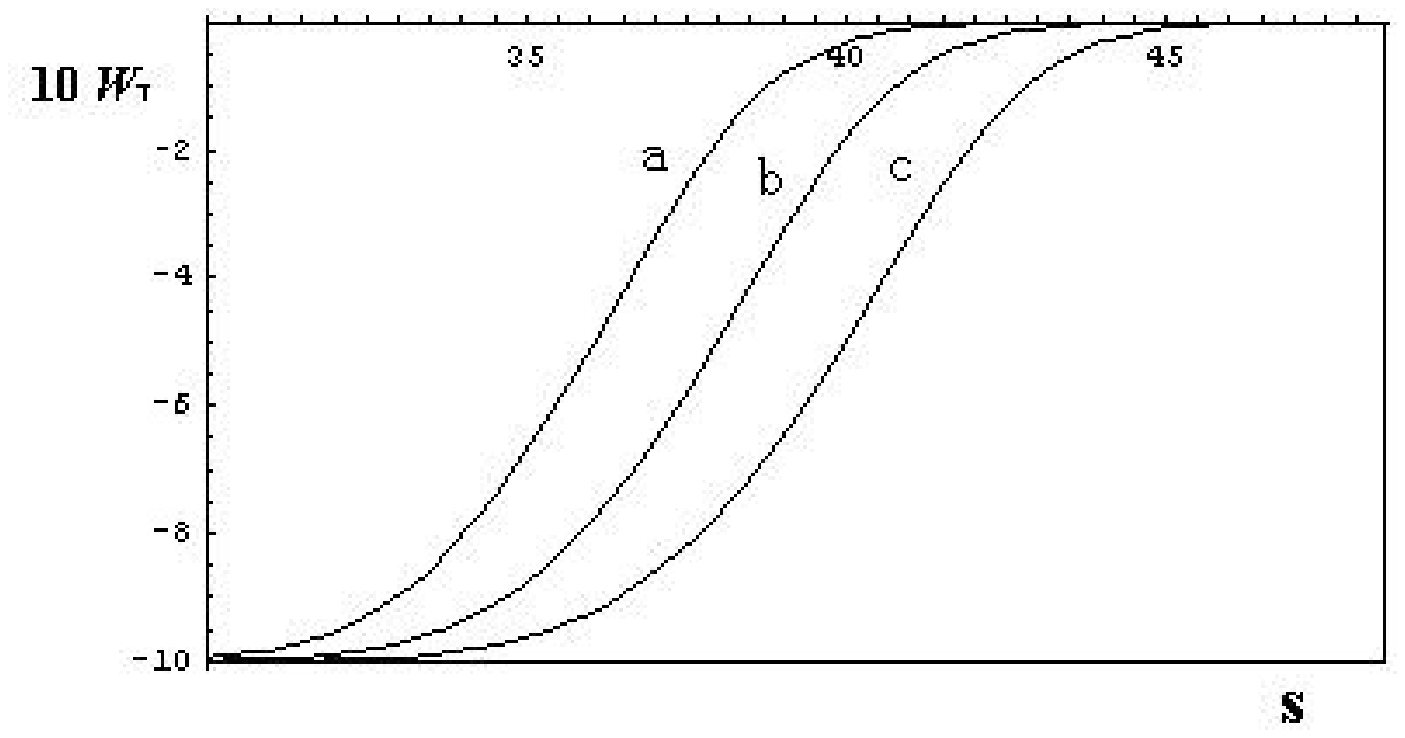,width=6in,height=4in}}
\caption{The plot of equation of state $w$ against $s$, curve a, b
and c correspond to the cases of $\beta=0.9, 1.0, 1.1$
respectively.}
\end{figure}

From Fig.1, one can find that the tachyon field increases steadily
with respect to $s$. From Fig.2, one knows that the variation of
the field will approach an asymptotic value. The Fig.3 shows that
the equation of state $w_T$ is -1 at the beginning and then
increases to 0 as time evolves. If one consider the tachyon as
dark energy that accelerates the expansion of current universe,
then this acceleration will eventually stop and then the universe
will recover its decelerating expansion.

\vspace{0.4cm} \noindent\textbf{IV. Discussion and Conclusion}
 \vspace{0.4cm}

In this paper, we derive the relation between the equation of
state of rolling tachyon of string theory and the observable
coordinate distance in term of red-shift $z$ of SNe Ia. the
prerequisite of such a relation is that we consider the tachyon as
the dark energy that dominates the universe and accelerates its
expansion at recent epoch. In some earlier works\cite{turner2},
such a possibility has been pointed out and the theoretical
predication and the reconstructed results have been compared with
the simulated SNe Ia data, which is an important and interesting
topic.

The importance of the reconstruction equations lies in that it may
serve as the intersection between theory and observations. We
analyzed numerically the dynamical evolution of the equation of
state of tachyon in the potential derived from string theory, and
found that it will increase steadily with time, which implies that
the expansion of the universe will eventually slow down and become
decelerating. This could be compared with the observed SNe Ia data
through the reconstruction equation in the former part of this
paper. It is worth noting that the evolution of the equation of
state in this paper is under the initial condition that the
tachyon is rolling down from the top of the potential with 0
initial speed. The initial condition could be choose otherwise so
that the equation of state evolves toward -1, that is, the tachyon
rolls up to the top of potential although this is not very
natural. In some other effective potentials, one may find
different evolution of the $w_T$, which can be compared with the
reconstructed evolution of $w_T$.

On the other hand, the tachyon here we considered is resulted from
string theory, while there are also other possibilities. For
example, the tachyon inflation has also been considered for
phenomenological potentials that are not derived from string
theory\cite{Feinstein}. Such models are related to
phenomenological " k-inflation"\cite{picon}. Clearly, this work
can also be further generalized to these kinds of phenomenological
theories.

\vspace{0.8cm} \noindent ACKNOWLEDGMENTS

This work was partially supported by National Nature Science
Foundation of China, National Doctor Foundation of China under
Grant No. 1999025110, and Foundation of Shanghai Development for
Science and Technology 01JC1435.


\begin{thebibliography}{99}

\bibitem {turner1} M. S. Turner, Phys. Rev.
\textbf{D48}3502(1993); E. J. Copeland, E. W. Kolb, A. R. Liddle
and J. E. Lidsey, Phys. Rev. Lett. \textbf{71} 219(1993); Phys.
Rev. \textbf{D48}2529(1993); Phys. Rev. \textbf{D49}1840(1994); A.
R. Liddle and M. S. Turner Phys. Rev. \textbf{D50} 758(1994)
\bibitem {turner2} D. Huterer and M. S. Turner, Phys. Rev.
\textbf{D60} 081301(1999)
\bibitem {steinhardt} B. Ratra and P. J. Peebles, Phys. Rev. {\bf
D37}3406(1988); R. R. Caldwell, R. Dave and P. J. Steinhardt,
Phys. Rev. Lett. {\bf 80}1582(1998); P. J. Steinhardt, L . Wang
and I . Zlatev, Phys. Rev. {\bf D59}123504(1999); I. Zlatev, L.
Wang and P. J. Steinhardt, Phys. Rev. Lett. {\bf 82}896(1999); K.
Coble, S. Dodelson, J. Frieman, Phys. Rev. {\bf D55}1851(1997).
\bibitem {Sen}A. Sen  , hep-th/0203211\\A. Sen , hep-th/0203265\\A. Sen , hep-th/0204143
\bibitem {Gibbons} G. W. Gibbons, hep-th/0204008
\bibitem {Fairbairn} M. Fairbairn and M. H. G. Tytgat ,hep-th/0204070
\bibitem {Feinstein} A. Feinstein , hep-th/0204140
\bibitem {Mukohyama} S. Mukohyama , hep-th/0204084


\bibitem {Gerasimov}Anton A. Gerasimov, Samson L. Shatashvili, JHEP \textbf{0010} 034(2000)
\bibitem {Mazumdar}A. Mazumdar, S. Panda, A. Prez-Lorenzana, Nucl.Phys. \textbf{B614} 101-116(2001)
\bibitem {Padmanabhan}T. Padmanabhan, Phys.Rev. \textbf{D66} 021301(2002)
\bibitem {Frolov}A. Frolov, L. Kofman and A. Starobinsky, hep-th/0204187
\bibitem {Minahan}Joseph A. Minahan, hep-th/0205098
\bibitem {Kim}H. Kim, hep-th/0204191
\bibitem {Choudhury}D. Choudhury, D. Ghoshal, D. P. Jatkar , S. Panda , hep-th/0204204

\bibitem {shiu}G. Shiu, I. Wasserman, hep-th/0205003

\bibitem {Linde}L. Kofman and A. Linde, hep-th/0205121

\bibitem {Maggiore}M. Maggiore, hep-th/0205014
\bibitem {Sugimoto}S. Sugimoto, S. Terashima, hep-th/0205085

\bibitem {Benaoum}H.B. Benaoum, hep-th/0205140
\bibitem {Sami}M. Sami, hep-th/0205146;  M. Sami, P. Chingangbam and T. Qureshi,
, hep-th/0205179
\bibitem {Li1} X. Z. Li, D. J. Liu and J.G. Hao, hep-th/02071460; J. Hwang, H. Noh
,hep-th/0206100; A. Ishida, S. Uehara, hep-th/0206102; T. Sarkar,
hep-th/0206109; H. Hata, S. Moriyama, hep-th/0206208; T. Mehen, B.
Wecht, hep-th/0206212; K. Ohta, T. Yokono, hep-th/0207004; N.
Moeller, B. Zwiebach, hep-th/0207107; G. Shiu, S.-H. Henry Tye, I.
Wasserman, hep-th/0207119; J. M. Cline, H. Firouzjahi, P.
Martineau,hep-th/0207156; A. Buchel, P. Langfelder, J. Walcher,
hep-th/0207235;G. Felder, L. Kofman , A. Starobinsky,
hep-th/0208019; A.V. Yurov, hep-th/0208075; S. Mukohyama,
hep-th/0208094; M. C. Bento, O. Bertolami, A.A. Sen,
hep-th/0208124; W. Taylor, hep-th/0208149;T. Okuda, S. Sugimoto,
hep-th/0208196;N.D. Lambert, I. Sachs, hep-th/0208217; G. Gibbons,
K. Hashimoto, P. Yi, hep-th/0209034;
\bibitem {Li} X. Z. Li, J. G. Hao and D. J. Liu, "Can quintessence
be the rolling tachyon?", hep-th/0204252
\bibitem {gerasimov} A. A. Gerasimov and S. L. Shatashvili, JHEP
\textbf{0010} 034(2000)
\bibitem {kutasov} D. Kutasov, M. Mari\~{n}o and G. W. Moore,
JHEP, \textbf{0010} 045(2000)
\bibitem {picon} C. Armendariz Pic\'{o}n, T. Damour and V.
Mukhanov, Phys. Lett. \textbf{B458}, 209(1999)


\end{thebibliography}
\end{document}